# High Pressure Studies on Fe-Pnictide Superconductors

C. W. Chu[1,2,3] and B. Lorenz[1]

[1]Department of Physics and TCSUH, University of Houston, Houston, TX 77204-5002, USA

[2]Lawrence Berkeley National Laboratory, 1 Cyclotron Road, Berkeley, CA 94720, USA

[3]Hong Kong University of Science and Technology, Hong Kong, China

**Corresponding Author**

C. W. Chu

Texas Center for Superconductivity

University of Houston

202 Houston Science Center

Houston TX  77204-5002

PHONE: 713-743-8222

FAX: 713-743-8201

e-mail: cwchu@uh.edu

**Abstract**

A review of high-pressure studies on Fe-pnictide superconductors is given. The pressure effects on the magnetic and superconducting transitions are discussed for different classes of doped and undoped FeAs-compounds, ROFeAs (R = rare earth), AeFe$_2$As$_2$ (Ae = Ca, Sr, Ba), and AFeAs (A = Li, Na). Pressure tends to decrease the magnetic transition temperature in the undoped or only slightly doped compounds. The superconducting $T_c$ increases with pressure for underdoped FeAs-pnictides, remains approximately constant for optimal doping, and decreases linearly in the overdoped range. The undoped LaOFeAs and $AeFe_2As_2$ become superconducting under pressure although nonhydrostatic pressure conditions seem to play a role in $CaFe_2As_2$. The superconductivity in the (undoped) AFeAs is explained as a chemical pressure effect due to the volume contraction caused by the small ionic size of the A-elements. The binary FeSe shows the largest pressure coefficient of $T_c$ in the Se-deficient superconducting phase.

**1. Introduction**

Majorities of interactions in solids (and other states of matter) depend critically on the inter-atomic distance. The application of pressure changes the inter-atomic distance and modify the electronic and the phononic energy spectra of a solid without introducing any chemical complexity and keeping the physical complexity to a minimum. This is particularly so when the pressure is hydrostatic and the solid is isotropic. Therefore, high pressure techniques have been extensively used in recent years to explore the physical states of solids, to create new ground states in solids, to test theoretical models and to help develop new theories. For example, there are more naturally nonsuperconducting elements that have been turned into superconductors through the application of pressure in the last half-century than naturally occurring elemental superconductors [1]. It was the high pressure experiments on the A15 compound system to examine the correlation between superconductivity and lattice instabilities in the late 1970's and early 1980's that gave us the confidence that superconductivity could take place at temperatures above 30's K [2], in contrast to the then theoretical prediction [3]. Again, it was the high pressure experiments on cuprates that demonstrated that indeed a superconducting transition temperature ($T_c$) could be achieved above 50's K and suggested that the cuprate system warranted further exploration for novel physics and higher $T_c$, besides providing hints for the search for $T_c > $ 50's K [4]. In fact, the current record-$T_c$ of 164 K was achieved in $HgBa_2Ca_2Cu_3O_{9+\delta}$ (Hg1223) with an ambient $T_c$ of 134 K only under a pressure ~ 30 GPa [5].

The recent discovery by Hosono et al. of superconductivity in the electron-doped LaOFeAs (La1111) at 26 K [6] has ushered in a new era of high temperature superconductivity. The parent compound of undoped La1111 displays a layered structure of the tetragonal ZrCuSiAs type of space group P4/nmm (Fig. 1a). It consists of two layered substructures, namely, the alternating FeAs-layers and the LaO-layers. Each FeAs-layer is composed of a Fe-sheet sandwiched by two As-sheets with the Fe-atoms tetrahedrally coordinated to four As-atoms. Each LaO-layer consists of an O-sheet sandwiched between two La-sheets. FeAs-layers may be expected to form the active block for the charge carriers to flow, and the LaO-layers constitute the charge reservoir blocks that inject charge carriers into the active block to maintain the maximum layer-integrity of the FeAs-layers, as in the cuprate HTSs. It is a semimetal and undergoes an antiferromagnetic transition at $T_o$ ~ 150 K and becomes superconducting when charge carriers are introduced via electron-doping by partial substitution of O by F when $T_o$ is suppressed to 0 K. The report of superconductivity in electron-doped LaOFeAs has attracted great attention worldwide for three main reasons among others: 1) it is a new material system with a relatively high $T_c$; 2) it belongs to a layered transition-element-pnictide material class that is comprised of a large number of compounds [7] and thus gives great possibilities in the search for higher $T_c$; and 3) it contains a large concentration of the magnetic Fe, which is antithetic to superconductivity, and thus provides an avenue to examine the role of magnetism in high temperature superconductivity (HTS). It is therefore rather natural to hope that what high pressure has done to cuprate HTS can be repeated in Fe-pnictide superconductors in raising the $T_c$. In the ensuing months, many other superconducting Fe-pnictides have been discovered. They are grouped into four homologous series, 1111 (ROFeAs with R = rare-earth; AeFFeAs with Ae = alkaline earth), 122 (AeFe$_2$As$_2$; AFe$_2$As$_2$ with A = alkaline), 111 (AFeAs) and 011 (FeSe). When properly doped, the maximum $T_c$'s are ~ 55 K for 1111 [8], ~ 38 K for 122 [9, 10], and 12-25 K for 111 [11, 12], and ~ 9-14 K for 011 [13], respectively. Many high pressure experiments have been conducted and theoretical models proposed on these compounds. We shall first briefly review the results and discuss their implications here before presenting the details and discussions according to the different homologous series in later sections.

Immediately after the report of 26 K superconductivity, high pressure experiments on the electron-doped LaOFeAs were carried out [14]. The reasonably high rate of $T_c$-enhancement by pressure of +1.2 K/GPa coupled with the rapid $T_c$-rise to 55 K by replacement of La observed in electron-doped La1111 by rare-earth elements of smaller ionic radii [8] gave scientists the optimism that a miracle similar to that for cuprate HTS was in the offing and the significant role of magnetism in HTS was about to unravel. Given the similarities between the Fe-pnictide and cuprate superconductors, the doping (x) dependence of the pressure effect on the $T_c$ of electron-doped Sm1111 (SmOFeAs) was soon investigated after the discovery [15]. Pressure was found to enhance $T_c$ for samples in the underdoped region, i.e. $dT_c/dx > 0$; to suppress $T_c$ for samples in the overdoped region where $dT_c/dx < x$; and to affect $T_c$ only slightly for the optimally doped sample, very different from the case of Hg1223 [5]. The observations suggest that the maximum $T_c$ of the R1111 may not exceed 50's K and that the lower-than-50's K observed then in R1111 may rise to the 50's K should the sample conditions be optimized. Unfortunately, the highest $T_c$ observed in superconducting Fe-pnictides to date has not exceeded 55 K, although the then lower-than-50 K-$T_c$ was raised to above 50 K. Pressure was also found to suppress the magnetic transition temperature $T_o$, consistent with the positive pressure effect on $T_c$ of the underdoped samples and with the doping effect on $T_o$ [15]. Later, more high pressure experiments were made on the R1111 phase compounds [16-22]. They demonstrated the close relationship between magnetism and superconductivity, the ability to raise the $T_c$ of close to optimally doped R1111 up to 45 K rapidly below ~ 5 GPa followed by a slow $T_c$-drop at higher pressures and the possibility to induce superconductivity in the undoped parent R1111 with a rapid $T_c$ up to ~ 25 K. Systematic high-pressure studies have also been carried out on the electron-doped R1111 obtained via O-deficiency through high pressure synthesis [20, 23, 24]. The $T_c$ of these O-deficient samples were observed to reach its maximum at ~ 50 K rapidly although at a slightly lower pressure of ~ 2 GPa similar to that observed in the F-doped R1111, re-enforcing the significant role of effective valence count of the anions or the carrier concentration independent of the details of doping.

In the process of searching for novel compounds that may have a higher $T_c$ than the doped R1111, $SrFe_2As_2$ and $BaFe_2As_2$ which were known to crystallize in a layered structure of the $ThCr_2Si_2$ with the I4/mmm space group (Fig. 1b) were revisited. Sr122 and Ba122 are semi-metals, and possess alternating FeAs-layers and antiferromagnetic transitions at ~ 200 K [25, 26] and 140 K [27], respectively, coinciding with a tetragonal-to-orthorhombic structural transition, similar to R1111. They become superconducting and form a new homologous series of $AeFe_2As_2$ (Ae122 with Ae = alkaline-earth elements) when hole-doped by partial replacement of Ae with alkaline elements = A [10, 28]. The large solubility range of $(Sr_{1-x}K_x)122$ make easy the formation of the full $T_c$-x phase diagram and thus enables the systematic study of the x-dependence of the pressure dependence on $T_c$. Indeed, it was found [29] that $dT_c/dP > 0$ for the hole-doped Sr122 samples in the region where $dT_c/dx > 0$; whereas $dT_c/dP < 0$ for samples in the region where $dT_c/dx < 0$, similar to the observation on the electron-doped Sm1111 [15]. Aside from what was concluded early, the observation shows a symmetry between electron- and hole-dopings consistent with previous doping results at ambient pressure [30]. Single crystals of members of the doped and undoped 122 have been successfully grown. They provide the unique opportunities to explore their physical properties in a more defined manner and make easier comparisons with theories. Among the interesting findings are: 1) the sensitivity of physical properties (especially the pressure induced superconductivity) of 122 single crystals to pressure hydrostaticity [31]; 2) pressure suppresses the magnetic transition temperature $T_o$ in all 122 compounds [29, 32-36]; 3) pressure induces superconductivity in undoped Sr122 and Ba122 up to ~ 38 K as by doping but only over a limited pressure range [17, 32, 34, 37, 38]; 4) conflicting reports exist concerning the pressure-induced superconductivity in Ca122 over a narrow pressure range around ~ 0.2-0.5 GPa [31, 39-41], leaving this case uncertain and hence raising similar questions about the nature of the pressure-induced superconductivity in Sr122 and Ba122; and 5) pressure generates a new phase designated as the "collapsed" tetragonal phase in Ca111 at low temperature at < 0.3 GPa, which is stable to higher temperature at high pressures [33, 42,43].

The observations show that: 1) magnetism and superconductivity are closely entangled similar to 1111; 2) special attention should be paid to the pressure conditions to extract meaningful data; and 3) the intriguing T-P phase diagram needs further study to extract possible important information concerning not just superconductivity but also the physico-chemical formation of the Fe-pnictides that may lead to higher $T_c$.

The 111 homologous series has only been recently discovered [11, 12, 44]. They (Li111 and Na111) crystallize in a layered tetragonal structure (P4/nmm) of the anti-PbFCl-type as shown in Fig. 1c, with the Li-ions occupying half of the Fe-sites in $Fe_2As$ ($Cu_2Sb$ type). The $Fe_2As_2$-layers consist of edge-sharing $FeAs_4$-tetrahedra. They can be derived from As capping of the Fe-square nets, above and below each center of the Fe-squares. Only one pressure experiment has been done on the Li111 [45]. It showed that pressure suppresses its $T_c$ linearly at a rate $dT_c/dP = -1.5$ K/GPa, a value close to that for the almost optimally electron-doped 1111 homologous series, suggesting that Li111 may exist in an electron-rich state, consistent with its thermoelectric data. It should be noted that Li111 are superconducting without external doping in contrast to theoretical predictions. The high pressure results may relieve the impasse.

The binary FeSe crystallizes in the tegragonal PbO-type structure α-FeAs phase (P4/mmm), which has become the simplest homologous 011 of the Fe-pnictide with an onset $T_c \sim 8$ K [13]. As shown in Fig. 1d, the compound displays the FeSe-layers similar to the FeAs-layers in the FeAs-pnictide superconductors. Under pressure, the $T_c$ rises rapidly to 27 K at 1.5 GPa at a rate $dT_c/dP \sim +9.1$ K/GPa, the highest among all Fe-pnictide superconductors [46]. The simple structure and the unusually large positive pressure effect on $T_c$ of the 011 phase may help unveil the mystery of superconductivity in Fe-pnicties and in cuprates in general.

Now we would like to summarize the high pressure experiments on the Fe-pnictides according to the homologous series to which they belong.

## 2. The R1111 Series (ROFeAs with R = rare-earth)

### 2.1. $R(O_{1-x}F_x)FeAs$

After the discovery of superconductivity at 26 K in electron doped $La(O_{1-x}F_x)FeAs$ [6], Lu et al. [14] were the first to examine the hydrostatic pressure effect on the $T_c$ of the polycrystalline optimally doped $La(O_{1-x}F_x)FeAs$ with nominal x = 0.11 and an ambient pressure onset $T_c$ = 26.3 K (determined resistively) and 23 K (magnetically). They employed the commercial pressure cell (Mcell10) to generate the hydrostatic pressure up to ~ 1 GPa and used the Quantum Design MPMS system to measure the magnetic susceptibility χ of the sample and the Mcell. They found that the $T_c$ increases with pressure at a rate $dT_c/dP$ ~ 1.2 K/GPa. The results together with the $T_c$-enhancement following the replacement of La by R of smaller radii [8] gave hope that the $T_c$ of the FeAs-layered compound system could be drastically raised.

Buoyed by the great news, we decided to determine the highest $T_c$ possible for the R1111 series and to determine the reason why $T_c$ is enhanced by hydrostatic pressure. In a similar approach adopted in the study of HTS cuprates, Lorenz et al. [15] decided to examine the pressure effect on $Sm(O_{1-x}F_x)FeAs$, that had the highest $T_c$ of 43 K at the time [8], with the nominal compositions x = 0.0, 0.05, 0.13 and 0.30, resistively, in a hydrostatic environment up to 1.7 GPa using the BeCu-clamp technique. The onset-$T_c$'s at ambient pressure of the samples studied are < 1.2 K, 24.7 K and 43.6 K for nominal x = 0.05, 0.13 and 0.3, respectively. The ρ-anomaly in the x = 0.05 sample indicative of the SDW transition is pushed toward lower temperature as shown in Fig. 2, representing the first direct evidence of the negative pressure effect on the SDW phase. The $T_c$ defined as the dρ/dT-peak temperature of the slightly electron-overdoped sample with a nominal x = 0.3 (the overdoped nature was later proven by reducing x to achieve a $T_c$ of 53 K) decreases with pressure at a rate of $dT_c/dP$ = - 2.3 K/GPa as shown in Fig. 3. Figure 4 shows the $T_c(P)$ for the electron-underdoped $Sm(O_{1-x}F_x)FeAs$ of a nominal x = 0.13 with a positive $dT_c/dP$ ~ +1.3 K/GPa below 0.8 GPa. Above 0.8 GPa, $T_c$ seems to saturate or, possibly, decrease toward higher pressure. This tendency of $T_c$, passing through a maximum with increasing pressure, has been confirmed for other underdoped $R(O_{1-x}F_x)FeAs$ compounds [16]. The observations can be summarized as: for electron doped $Sm(O_{1-x}F_x)FeAs$, $dT_c/dP$ > 0 when $dT_c/dx$ > 0 in the underdoped region and $dT_c/dP$ < 0 when $dT_c/dx$ < 0 in the overdoped region, similar to the HTS cuprates as far as the signs are concerned [47]. This implies that the main effect of pressure is to promote electron-doping in $Sm(O_{1-x}F_x)FeAs$, while it generates holes for the HTS cuprates. It should be noted that the suppression of the SDW transition observed should also contribute to the positive $dT_c/dP$ of the underdoped $Sm(O_{1-x}F_x)FeAs$. By referring back to the quadratic universal $T_c(x)$ relation for the hole-doped cuprates by Presland et al. [48], we have proposed a possible upper $T_c$-limit < 50's K for the R1111 FeAs-superconductors, independent of the R in R1111 and suggested to look for higher $T_c$ in the FeAs-layered compounds with more complex structures than that of the R1111 series [15, 47]. It was later shown theoretically by Nekrasov et al. [49] that R is indeed electronically isolated from the conducting electrons and plays no role in the superconductivity of Fe-pnictides. Unfortunately, our predicted upper $T_c$-limit remains valid to date; however, we hope the latter prediction will bear fruit soon.

At about the same time, Takahashi et al. [16] decided to raise the $T_c$ of La(O$_{1-x}$F$_x$)FeAs by pressure. They investigated the pressure effects on the resistivity ρ of the optimally electron-doped (with an x where $dT_c/dx \sim 0$) and the electron-underdoped (in a region where $dT_c/dx < 0$) La(O$_{1-x}$F$_x$)FeAs with x = 0.11 and 0.05. Their ambient onset $T_c$'s are 28 K and 9 K for the x = 0.11 and 0.05 samples, respectively. The pressure was generated in a BeCu piston-cylinder cell up to 3 GPa hydrostatically and up to 30 GPa in a diamond anvil cell quasi-hydrostatically. The ρ(T)'s of the x = 0.11 and 0.05 samples are shown in Figs. 5a and b at different pressures. The onset-$T_c$ of the x = 0.11 sample, as defined in Fig. 5a, increases steeply from 28 K to 43 K with pressure up to 3 GPa at a large initial rate $dT_c/dP \sim + 8$ K/GPa and then drops off slowly to 9 K at 30 GPa at a rate $dT_c/dP \sim - 1.4$ K/GPa as shown in Fig. 6. The onset-$T_c$ of the sample of x = 0.05 was also found to increase with pressure but at a slower rate of + 2.0 K/GPa below 3 GPa where the experiment stops (Fig. 7 and inset). The experiment did show that the $T_c$ of La(O$_{1-x}$F$_x$)FeAs with F-doping can be raised to a maximum of 43 K by the joint effects of chemical (F-doping) and pressure doping. The authors further pointed out that a large number of existing transition-metal-oxypnictes with a structure similar to LaOFeAs [7] may offer great opportunities for seeking superconductors of higher $T_c$ by proper doping. The experiment also shows the sensitivity of ρ and the superconducting transition of La(O$_{1-x}$F$_x$)FeAs to non-hydrostatic pressure effects as evident in Figs. 5a and b where ρ(T)'s at different pressures are displayed. Pressure inhomogeneity tends to broaden the superconducting transition and remove the zero-ρ state, making it difficult to define the onset $T_c$ at high pressure. This was further demonstrated later in the study of 122 Fe-pnictides where hydrostaticity may have a bearing on the nature of superconductivity induced by pressure as was observed decades ago in α-uranium.

It should be noted that Fratini et al. [50] suggested that internal chemical pressure may simulate the externally applied pressure effect on $T_c$ by analyzing and comparing the structural transitions of the undoped LaOFeAs and NdOFeAs. Later Wang et al. [51] carried out an iso-valent-doping experiment on LaOFe(As$_{1-x}$P$_x$) by partial replacement of As by P. Their XRD results show a reduction in volume as x increases. They found that LaOFe(As$_{1-x}$P$_x$) become superconducting with $T_c$ reaching 10.5 K at x = 0.4. Unfortunately, a higher $T_c$ was not realized.

The large pressure effect on the $T_c$ of the optimally doped La(O$_{1-x}$F$_x$)FeAs sample observed at low pressure appears to be in direct conflict with the universal $T_c(x)$-relation based on which our conjecture of an ultimate $T_c$ < 50's K was made. A zero $dT_c/dP$ is expected from the standard universal $T_c(x)$-relation. Such a prediction is true only when the rigid band model is true, i.e. the $T_c(x)$-relation is not altered by pressure in terms of the maximum $T_c$ and the optimal x where $T_c$ reaches its maximum. However, the $T_c(x)$-relation has been shown to undergo a general shift without changing the general features of the relation in the case of HgBa$_2$Ca$_{n-1}$Cu$_n$O$_{m+2n+2+\delta}$, whose maximum $T_c$'s and optimal x's are known to be shifted upward by pressure [52]. In other words, pressure changes the carrier concentration and the band simultaneously. Therefore, the positive $dT_c/dP$ for an optimally-doped La(O$_{1-x}$F$_x$)FeAs is understood in terms of this modified $T_c(x)$-model. However, the asymmetrical $T_c$-peak with respect to pressure is less obvious and whether it is a characteristic of the Fe-pnicitides is unclear.

Takahashi et al. [17], using the same pressure techniques, later extended their study to include different doping levels, x = 0 (undoped), 0.05 (underdoped), 0.11(optimally doped) and x=0.14 (overdoped) with their $T_c$'s ~ 0 K, 24, 29 and 20 K, respectively, at ambient pressure. The results are summarized in Fig. 8. Similar to their previous work on the x = 0.11 sample, all display rapid $T_c$-increase followed by a slower $T_c$-drop above ~ 5 GPa to 30 GPa and the rate is the highest of 8 K/GPa for the optimally doped sample. They also witnessed the undoped LaOFeAs to become superconducting at pressures above ~ 2 GPa. While the maximum $T_c$'s for the x = 0.11 and 0.14 samples are about the same at ~ 43 K, they are lower at ~ 29 K and 20 K for samples with x = 0.05 and 0. The data indicate that the nearly constant maximum $T_c$ close to the optimal doping at ambient pressure and the general nonlinear overall $T_c$-P behavior are consistent with the modified $T_c(x)$-model discussed above. In the same study, Takahashi et al. found that the $T_c$ of LaOFeP with an ambient pressure value of $T_c$ = 5.8 K rises steeply to 8.8 K at 1 GPa and then drops rapidly to ~ 7 K with further increasing pressure, similar to the LaOFeAs. The crystal structure of these samples was also studied by XRD at room temperature under pressures up to 10 GPa. They found a bulk modulus of ~ 70 GPa, similar to the cuprates, and a reduction in anisotropy by pressure but no pressure-induced structural change. Zocco et al. [19] found a qualitatively similar $T_c(P)$-behavior of $La(O_{1-x}F_x)FeAs$ with a more rapid drop of $T_c$ at high pressure. However, they found an exception of a continuous decrease of $T_c$ with pressure for the optimally doped $Ce(O_{1-x}F_x)FeAs$ (Fig. 9), showing that R in R1111 may play a role in their $T_c$-response to pressure, although there may still be uncertainty in the exact doping level. More examples with such behavior of negative effect on $T_c$ from low to high pressure were found, for instance, by Takeshita et al. [23] in another study on oxygen-deficient $NdO_{1-y}FeAs$. Garbarino et al. [18] measured the pressure effects on the $T_c$ and the lattice parameters of the optimally doped

La($O_{1-x}F_x$)FeAs. They found that $T_c$ decreases linearly with pressure up to ~ 20 GPa. Yi et al. [20] reported the similar dependence of $T_c$ on pressure for the nominally optimal-doped and overdoped La($O_{1-x}F_x$)FeAs. However, the exact values of x cited may be questionable in view of their relative $T_c$'s. Okada et al. [22] measured the $T_o$ and $T_c$ of undoped LOFeAs and found that both coexist in a wide range of pressures of ~ 1 -13 GPa. From the above reports, it is clear that the pressure effect on the $T_c$ of R1111 depends on R, the pressure hydrostaticity, and the doping level. The effect of pressure hydrostaticity above 2-3 GPa cannot be neglected and the uncertainty associated with the latter is large due to the complexity in sample preparation and characterization.

**2.2. $RO_{1-y}$FeAs**

Soon after electron-doping was found to induce superconductivity in LaOFeAs through F-doping, electron doping was also done successfully via introducing O-deficiency (y) to R1111 by high pressure synthesis technique by Ren et al. [53]. They achieved a $T_c$ as high as 54 K in these O-deficient samples $RO_{1-y}$FeAs (R=Sm, Nd, Gd). Takeshita et al. [23] examined resistively the pressure effect on the $T_c$ of $NdO_{1-y}$FeAs with a nominal 1-y = 0.6, 0.8 and 0.85 with an ambient onset $T_c$ = 54 K, 41 K and 0 K, respectively, up to 18 GPa. They found that the onset-$T_c$'s of the superconducting samples decrease continuously with pressure-increase at a rate of $dT_c/dP$ ~ - 2 K/GPa, as shown in Fig. 10. For 1-y = 0.85, superconductivity appears after the signature of the SDW transition is suppressed by pressures above 5 GPa with an onset $T_c$ that continues to rise to ~ 20 K at 16 GPa. Later, Yi et al. [24] examined the $\rho(T)$ of Sm- and $NdO_{1-y}$FeAs with 1-y = 0.85 under pressures up to ~ 7GPa. Again, $T_c$'s for both samples decrease monotonically with pressure-increase as displayed in Fig. 11. The $SmO_{1-y}$FeAs sample was also investigated under a hydrostatic pressure up to 1 GPa and showed the same $T_c(P)$ behavior. No $T_c$-maximum with pressure has been observed. Whether its absence is characteristic of $RO_{1-y}FeAs$ or due to the lack of detailed studies remains to be determined. If the former is true, the local atomic structure may play an important role in the superconductivity of As-pnictides.

## 3. The 122 Series [$(Ae_{1-x}A_x)Fe_2As_2$ with Ae = alkaline earth, $Ae(Fe_{1-x}Co_x)_2As_2$, and $AeFe_2As_2$]

### 3.1. Hole-doped $(Ae_{1-x}A_x)Fe_2As_2$

$SrFe_2As_2$ and $BaFe_2As_2$ were known to crystallize in a layered structure of the $ThCr_2Si_2$ with the I4/mmm space group [54]. They are semi-metals, and possess alternating FeAs-layers and antiferromagnetic transitions at ~ 200 K (Sr) and 140 K (Ba), coinciding with a tetragonal-to-orthorhombic structural transition, similar to R1111. They become superconducting and form a new homologous series of 122 when hole doped, i.e. by partial replacement of Ae with alkaline elements = A, $(Ae_{1-x}A_x)Fe_2As_2$ [9, 10, 25, 28]. The large solubility range of $(Sr_{1-x}K_x)Fe_2As_2$ makes it possible to form a complete $T_c$-x phase diagram at ambient pressure, as shown in Fig. 12 where $T_o$ represents the SDW transition temperature. The maximum $T_c$ = 38 K takes place at x ~ 0.4 - 0.5. Possible coexistence of SDW and the superconducting states is apparent between x ~ 0.2 and 0.3. The phase diagram makes possible the systematic study of the x-dependence of the pressure coefficient of $T_c$ of the 122 series. It should be noted that this was the first complete $T_c(x)$-phase diagram of Fe-pnictides available at the time. We have, therefore, measured $\rho(T)$ and $\chi_{ac}(T)$ under hydrostatic pressures up to 1.7 GPa, for different x's to cover the three major regions of hole-doping [10, 15, 29]. The $T_c(P)$ for samples with x = 0.2 (underdoped), 0.4 (~ optimally-doped) and 0.7 (overdoped) are given in Fig. 13. It is unambiguously shown that $dT_c/dP$ is positive for the underdoped sample, close to zero for the optimally doped sample and negative for the overdoped sample, in agreement with the prediction of the universal $T_c(x)$-relation proposed for the hole-doped cuprate HTSs as discussed in section 2.1. The observation can be summarized as: $dT_c/dP > 0$ in the region where $dT_c/dx > 0$; $dT_c/dP \sim 0$ where $dT_c/dx \sim 0$; and $dT_c/dP < 0$ where $dT_c/dx < 0$, suggesting that pressure promotes holes. The pressure study on electron-doped $Sm(O_{1-x}F_x)FeAs$ suggests that pressure generates electrons. Based on these two studies, one may conclude that symmetry between the electron- and hole-dopings exists in Fe-pnictide superconductors, in agreement with the earlier observation by Wen et al. [30]. It should be noted that pressure suppresses the SDW transition in the hole-doped $(Sr_{1-x}K_x)Fe_2As_2$ (Fig. 13), just as in the electron-doped $Sm(O_{1-x}F_x)FeAs$.

A single crystal of the isoelectronic 122 sample $(Ba_{1-x}K_x)Fe_2As_2$ with $x = 0.45$ and a $T_c = 30$ K was grown and examined under hydrostatic pressure using a BeCu-clamp [36]. $T_c$ decreases with pressure at a rather small rate of $-0.21$ K/GPa, which puts the compound into the slightly overdoped region, consistent with our conclusions drawn from the $(Sr_{1-x}K_x)122$ system.

## 3.2. Electron-doped $Ae(Fe_{1-x}Co_x)_2As_2$

The symmetry of the superconducting and magnetic phase diagram with respect to electron and hole doping, first observed in the R1111 series of compounds, was also verified for the Ae122 class of pnictides. Whereas replacing alkaline earth (Ae) with alkaline metal (A) ions introduces holes into the bands of the undoped $AeFe_2As_2$ the substitution of Fe with Co corresponds to electron doping. Superconductivity with a maximum $T_c$ of 22 K was discovered in $Ba(Fe_{1-x}Co_x)_2As_2$ (x=0.1) [55] and a complete phase diagram, similar to that shown in Fig. 12 for hole-doped $(K_xSr_{1-x})Fe_2As_2$, was revealed [56]. The effects of hydrostatic pressure up to 2.5 GPa on the SDW and superconducting transitions have been investigated by Ahilan et al. [57] for an underdoped (x=0.04) and an optimally doped (x=0.1) sample of $Ba(Fe_{1-x}Co_x)_2As_2$. The results are surprisingly similar to our results obtained for the hole-doped $(K_xSr_{1-x})Fe_2As_2$. The underdoped sample (x=0.04) shows a suppression of the SDW transition and an increase of the superconducting $T_c$ whereas the $T_c$ of the optimally doped compound (x=0.1) is nearly independent of pressure. This remarkable similarity of the pressure effects in both, hole and electron doped Ae122 pnictides, shows the electron-hole symmetry in the pnictide superconductors, similar to the high-temperature cuprate superconductors. The structural anisotropy of the layered pnictides raises the question about possible differences between hydrostatic and unaxial pressure effects. Based on thermal expansion and heat capacity measurements of single crystals of the underdoped $Ba(Fe_{1-x}Co_x)_2As_2$ (x=0.04), Hardy et al. [58] derived the uniaxial pressure coefficients of $T_c$ for compression along the a- and c-axes. The pressure coefficients are quite large and very anisotropic, $dT_c/dp_a$=3.1 K/GPa and $dT_c/dp_c$=-7.0 K/GPa, indicating that $T_c$ of $Ba(Fe_{1-x}Co_x)_2As_2$ strongly depends on the c/a ratio, similar to other layered superconductors (for example: $La_{2-x}Sr_xCuO_4$ [59]).

### 3.3. Undoped $AeFe_2As_2$

The Ae122 homologous series comprise three members, namely, $CaFe_2As_2$, $SrFe_2As_2$ and $BaFe_2As_2$ [54]. These undoped parent compounds are semi-metallic and have the tetragonal layered crystal structure at room temperature. Upon cooling, they undergo a tetragonal (T) to orthorhombic (O) structural transition at $T_o$ coinciding with an antiferromagnetic transition [26, 27]. Similar to the R1111 homologous series, pressure is found to suppress the magnetic order [32, 35, 41]. However, the occurrence of superconductivity in Ae122 under pressure appears to be more complicated than in R1111. For instance, some reported the observation of pressure-induced superconductivity in all three Ae122 members in different narrow pressure-ranges, while others raised serious doubt of its existence in Ca122 and yet others questioned about the nature of the pressure-induced superconductivity in all three Ae122 compounds. These divergent views indeed are extremely puzzling, given the fact that bulk superconductivity has been reported in the chemically doped samples of all members of the Ae122 series and the great similarities among the Ae122 members. A moderate pressure below 1 GPa apparently can tune into different ground states for Ca122 in its T-P phase space, providing new opportunities to resolve the above puzzles and to unveil the interplay between superconductivity, magnetism, structural instabilities and local atomic structures. The success in growing single crystals of Ae122, in contrast to the difficulty encountered for R1111, makes the study of Ae122 timely. We shall present representative high pressure data below:

### 3.3.1. $CaFe_2As_2$

Torikachvili et al. [41] have measured the $\rho(T)$ of single crystal $CaFe_2As_2$ under hydrostatic pressures up to 2 GPa using the BeCu-clamp technique. The $\rho(T)$'s at different pressures are shown in Fig. 14. At ambient pressure, $\rho(T)$ decreases with temperature initially, rises abruptly at $T_o \sim 170$ K indicating the onset of a first-order structural (T-O) transition coinciding with the antiferromagnetic transition, and continues to drop afterward. No superconductivity was detected to ~ 0.6 K. Under small pressures up to 0.35 GPa the structural transition is broadened and $T_o$ pushed down to 130 K continuously (blue dots in Fig. 15). However, at pressures > 0.35 GPa, a new $\rho$-anomaly appears with a clear $\rho$-drop on cooling over a certain temperature-range, signaling the entrance to a new phase transition (cT). The anomaly becomes clearer with pressure as its temperature is shifted to higher temperature (red dots in Fig. 14). The new transition (T-cT) is found thermally hysteretic like the structural transition at $T_o$. A ρ-drop at ~ 12 K indicative of the beginning of a superconducting transition was evident at 0.23 GPa and zero-ρ was also achieved but only within a narrow pressure region between 0.35 and 0.55 GPa. With the transition temperatures defined in Fig. 14 and the ρ(T)-behavior, the authors constructed the first T-P phase diagram shown in Fig. 15.

To clarify the nature of different phases in the T-P diagram, many studies have been made. Park et al. [40] determined the $H_{c2}(0)$ of the superconducting state at 0.69 GPa to be 10-14 T. Kreyssig et al. [42] found via neutron scattering that $CaFe_2As_2$ undergoes a transition from the nonmagnetic T-phase to the antiferromagnetic O-phase at low pressure. At high pressures, it goes from the nonmagnetic T phase to the nonmagnetic cT-phase where the Fe-moment is quenched. The cT-phase is known as the collapsed-phase because of the drastic reduction of 9.5% of its c-lattice parameter and 11% of the c/a ratio in comparison with the O-phase at low temperature. The superconducting state is supposed to occur in the nonmagnetic cT-phase, in contrast to the original thinking that the superconducting state exists in the magnetic O-phase. Some interesting suggestions were made concerning the importance of As-Fe-As bond angle and magnetic fluctuations to superconductivity in As-pnictides.

Goldman et al. [33] examined single crystals of $CaFe_2As_2$ by neutron and high-energy x-ray diffraction to identify the phase lines for transitions between the high-temperature ambient-pressure tetragonal (T), the antiferromagnetic orthorhombic (O) and the nonmagnetic collapsed tetragonal (cT) phases as shown in Fig. 16. No magnetic ordering was found in the cT-phase.

Both T-O and T-cT transitions are hysteretic during isothermal pressure-cycling and isobaric temperature-cycling, although the effect for the T-O transition is much smaller than for the T-cT transition as represented by the shaded area in Fig. 16. The effect is beautifully demonstrated in Fig. 17. At room temperature, associated with the T-cT transition are a 10% reduction of the c-axis and a 2% expansion of the in-plane lattice parameters. Superconductivity was detected only near the T-cT transition in the hysteretic area. A question concerning its nature was raised. The loss of magnetism was attributed to the depletion of the Fe 3d density of states at the Fermi surface as was suggested by band-calculations. Yu et al. [31] performed a careful magnetic susceptibility experiment under hydrostatic pressure conditions using He as the pressure medium and they detected no sign of superconductivity. They therefore concluded that neither the O nor the cT-phase is superconducting. They further attributed the occasional observations of superconductivity near the T-cT transition to possible intermediate phases generated by non-hydrostatic pressure conditions in domain walls of phase separated regions arising from the large anisotropic change crossing the O-cT transition line. Pratt et al. [43] recently did an inelastic neutron scattering study on $CaFe_2As_2$ under hydrostatic pressures and found the antiferromagnetic fluctuations in all three phases are suppressed or quenched completely in the cT-phase within the experimental resolution. They tried to relate the suppressed magnetic fluctuations with the absence of superconductivity and proposed a significant role of the magnetic fluctuations in superconductivity of Fe-pnictides.

### 3.3.2. $SrFe_2As_2$

The emergence of superconductivity above a critical pressure in the undoped $SrFe_2As_2$ has been reported by different groups. Though all report the pressure-induced superconductivity, the reported data differ quantitatively with respect to the critical pressure, the maximum $T_c$, and the details of the T-P phase diagram. Alireza et al. [37], utilizing dc magnetization measurements of a single crystal of $SrFe_2As_2$ mounted in a diamond anvil cell, have shown a sudden onset of superconductivity at about 2.5 GPa with a $T_c \sim 27$ K. $T_c$ did decrease to below 20 K with further increasing pressure to 5.2 GPa. The results of P-induced superconductivity had been confirmed by Torkachvili et al. [36] with a similar critical pressure for the onset of superconductivity. However, the superconducting state ceases to exist above a pressure of 3.7 GPa, in contrast to the earlier work. Different high-pressure techniques have been employed by Takahashi et al. [17, 38]. The superconducting phase of $SrFe_2As_2$ was reported to arise at pressures below 2 GPa and to extend to pressures as high as 14 GPa with a maximum $T_c$ of 34 K, close to the maximum $T_c$ achieved at ambient pressure in the hole-doped system $(Sr_{1-x}K_x)Fe_2As_2$ [10]. Kotegawa et al. [34] conducted careful resistivity measurements of $SrFe_2As_2$ single crystals under pressure and detected the zero-resistance superconducting state at pressures above 3.5 GPa with a maximum $T_c$ of 34 K. The SDW transition temperature decreased with P and the AFM/orthorhombic phase seemed to be suppressed by pressure above 3.7 GPa. The T-P phase diagram of $SrFe_2As_2$ showing the superconducting and magnetic phase boundaries is shown in Fig. 18.
The significant differences in the critical pressures for the onset and, at higher pressure, suppression of the superconducting phase in $SrFe_2As_2$ show that the P-induced superconductivity is very sensitive with respect to sample quality, pressure hydrostaticity, method of measurement, and the selected criterion defining $T_c$. Similar to $CaFe_2As_2$ (Section 3.3.1) the superconducting state appears to be fragile and competing with other magnetic or non-magnetic states.

### 3.3.3. $BaFe_2As_2$

The onset of superconductivity in BaFe$_2$As$_2$ above a critical pressure of 2 to 2.5 GPa and a maximum T$_c$ ~ 29 K was consistently reported by two groups [32, 37]. The superconducting phase appears to extend to rather high pressures (> 6 GPa [37] and > 13 GPa [32]). Similar to the two other members of the series of Ae122, the application of external pressure does decrease the SDW transition temperature from its ambient pressure value of 140 K.

The T-P phase diagrams for BaFe$_2$As$_2$ and SrFe$_2$As$_2$ reveal an interesting similarity with the T-x phase diagrams of the electron or hole doped AeFe$_2$As$_2$. Starting from the undoped (ambient pressure) non-superconducting but magnetic compounds the introduction of electron or hole carriers (application of pressure) first reduces the SDW transition temperature and eventually introduces the superconducting state above a critical doping (pressure). The possible coexistence of the SDW and superconducting states in some doping (pressure) range is of particular interest. For the hole doped (Sr$_{1-x}$K$_x$)Fe$_2$As$_2$ it was shown that the typical signature of the SDW transition could be detected above the superconducting T$_c$ in a narrow doping range [29]. Similar conclusions have been drawn for (K$_x$Ba$_{1-x}$)Fe$_2$As$_2$ [60] and for the electron-doped system Ba(Fe$_{1-x}$Co$_{1-x}$)$_2$As$_2$ [61]. The coexistence of SDW and superconducting phases in the T-P phase diagrams of AeFe$_2$As$_2$, however, has yet to be explored. Some experimental investigations show the loss of the characteristic anomaly indicating the SDW transition with the abrupt onset of superconductivity close to the critical pressure suggesting a sudden suppression of the magnetic phase [34-36] whereas others are in favor of the coexistence of both orders in a broad pressure range [32, 38]. Further and more detailed work is needed to resolve these issues.

**4. Pressure Effects in the A111 System (LiFeAs)**

Superconductivity found in LiFeAs [11, 12] at 18 K, and later in NaFeAs [44] at 12 K, seems to be puzzling since in the stoichiometric 111 composition these compounds are expected have a similar electronic structure in the $Fe_2As_2$-layers (average charge: 1 electron per FeAs) as the undoped R1111 and Ae122 pnictides. No indication of a SDW transition could be found in different physical quantities and Mössbauer measurements have clearly shown that long range magnetic order does not exist in LiFeAs [62].

We have therefore conducted high-pressure measurements of the superconducting $T_c$ of LiFeAs in order to gain more insight into the complex physics of the A111 pnictides [45]. The superconducting $T_c$ was found to decrease linearly with pressure at a rate of -1.5 K/GPa. This significant pressure effect is difficult to understand. The large and linear decrease of $T_c$ is comparable with the pressure coefficients of overdoped R1111 or Ae122 pnictides, however, the formal charge balance (1 electron per FeAs) locates LiFeAs close to the undoped pnictides. The resolution may lie in the short lattice constants of LiFeAs due to the small size of the Li ion. LiFeAs indeed is among the pnictides with the smallest volume and it can therefore be considered a "high pressure" analogue of the undoped FeAs-compounds (the lattice contraction caused by the smaller ions is frequently called "chemical pressure"). This could explain the superconductivity with a relatively high $T_c$ in LiFeAs as well as NaFeAs since superconductivity is also induced by (physical) pressure in the undoped LaOFeAs and $AeFe_2As_2$, which all show a negative pressure coefficient towards higher pressure after passing through the maximum of $T_c$. This conjecture has received further support recently by investigations of the effect of isovalent substitution of P for As, as for example in undoped $EuFe_2(As_{0.7}P_{0.3})_2$ [63] and $BaFe_2(As_{1-x}P_x)_2$ [64]. The replacement of As by the smaller P ion reduces the lattice parameters and the volume (chemical pressure). The SDW phase is completely suppressed and superconductivity arises with $T_c$'s up to 26 K (Eu) and 30 K (Ba). With these results we can expect a further increase of $T_c$ of LiFeAs if the lattice can be expanded (negative pressure). Thin films of LiFeAs on a substrate with an appropriate lattice mismatch (tensile strain) are therefore expected to exhibit higher $T_c$'s and improved superconducting properties.

**5. The 011 Series, $FeSe_x$**

The search for new layered compounds similar and alternative to the FeAs-superconductors has led Hsu et al. to investigate the physical properties of the PbO-type structure α-FeSe [13]. The structure of FeSe is equivalent to the structure of LiFeAs with all Li ions removed (Fig. 1d). Superconductivity with a transition temperature of 8 K was revealed in Se-deficient samples. The $T_c$ of $FeSe_x$ could be increased to 27 K (onset of the resistivity drop) by applying pressure up to 1.5 GPa [46].

Li et al. [65] conducted a detailed investigation of the effect of pressure on the magnetic and superconducting states of α-$FeSe_x$ for x=0.8 and 0.88. Magnetic measurements show three distinct anomalies upon increase in temperature, a diamagnetic signal below the superconducting transition at 8 K, a rapid rise of magnetization at about 78 K, and a magnetization decrease at 106 K. The latter two transitions have been ascribed to an antiferromagnetic and a ferromagnetic transition, respectively. However, high-resolution X-ray diffraction measurements have shown that these two phase transitions are also structural transformations associated with changes of the lattice symmetry [13, 66]. Under external pressure all three phase-transition temperatures are shifted to higher temperatures, which was interpreted as a sign of coexistence of superconductivity with magnetism [65]. However, the true nature of the magnetic anomalies has yet to be revealed.

## 6. Conclusion

High pressure investigations of iron pnictide superconductors have contributed significantly to a better understanding of this class of new superconducting compounds. Initial hopes that high pressure might increase the critical temperatures beyond the maximum ambient pressure values of 55 K (R1111) have not yet been confirmed. Instead, in the low-pressure range it was shown for quite different compounds and doping (R1111, hole and electron doped Ae122, A111) that the pressure shift of the superconducting $T_c$ strongly depends on the doping state. Pressure does enhance $T_c$ in the underdoped state, $dT_c/dp > 0$, $T_c$ is nearly independent of pressure for optimally doped compounds, $dT_c/dp \sim 0$, and pressure suppresses $T_c$ in the overdoped state, $dT_c/dp < 0$, revealing an interesting analogy to the copper oxide high-temperature superconductors.

Close to the undoped state of R1111 and Ae122 a spin density wave phase is observed below 200 K. The magnetic transition temperature is suppressed by pressure as well as by doping while the superconducting $T_c$ is enhanced, demonstrating the close relationship of superconductivity and magnetism. The undoped La1111 and Ae122 pnictides become superconducting under pressure. However, the details of the temperature-pressure phase diagram differ significantly between different compounds and the P-induced superconductivity in Ca122 has been questioned. The observation of superconductivity in the A111 series (LiFeAs, NaFeAs) in the undoped state has been a puzzle. The strong and linear decrease of $T_c$ in LiFeAs and the small volume of this compound in the ambient pressure state show that the A111 compounds are the high-pressure analogue of the undoped La1111 or Ae122 pnictides.

The binary FeSe exhibits a similar layered structure as the FeAs-pnictide superconductors. The Se-deficient compounds are superconducting below 8 K. They show the highest positive pressure coefficient, $dT_c/dp \sim 9.1$ K/GPa, among all Fe-pnictide superconductors. The maximum $T_c$ in FeSe upon further increasing pressure has yet to be revealed.

**Acknowledgments**

This work is supported in part by the T.L.L. Temple Foundation, the John J. and Rebecca Moores Endowment, the State of Texas through TCSUH, the U.S. Air Force Office of Scientific Research, and the LBNL through the U.S. Department of Energy.

**Figure Captions**

Fig. 1: Structure of (a) 1111, (b) 122, (c) 111, and (d) 011 homologous series of Fe-pnictides [(d) from Ref. 13].
Fig. 2: Negative pressure effect on the SDW temperature $T_o$ of Sm1111.
Fig. 3: Pressure effect on the $T_c$ of the electron-overdoped Sm1111 with a nominal x=0.3.
Fig. 4: Pressure effect on the $T_c$ of the electron-underdoped Sm1111 with a nominal x=0.13.
Fig. 5: $\rho(T)$ of the La1111 with x=0.11 at different pressures (from Ref. 16).
Fig. 6: $T_c(P)$ of La1111 with x=0.11 (from Ref. 16).
Fig. 7: $\rho(T)$ and $T_c(P)$ of La1111 with x=0.05 (from Ref. 16).
Fig. 8: $T_c(P)$ of La1111 with x=0.00, 0.05, 0.11, and 0.14 (from Ref. 17).
Fig. 9: $T_c(P)$ of Ce1111 with x=0.12 (from Ref. 19).

Fig. 10: $T_c(P)$ of $NdO_{1-y}FeAs$ (from Ref. 23).
Fig. 11: $T_c(P)$ of Nd- and $SmO_{1-y}FeAs$ (from Ref. 24).
Fig. 12: $T_c$ vs. x and $T_o$ vs. x for $(Sr_{1-x}K_x)Fe_2As_2$
Fig. 13: $T_c(P)$ for hole-doped $(Sr_{1-x}K_x)Fe_2As_2$.
Fig. 14: $\rho(T)$ at different pressure for undoped Ca122 single crystals (from Ref. 41).
Fig. 15: T-P phase diagram for $T_o$, $T_{o2}$ and $T_c$ of undoped Ca122 single crystals (from Ref. 41).
Fig. 16: The T, O and cT phase boundaries of Ca122 (from Ref. 33).
Fig. 17: $\rho(T)$ close to the T-O and O-cT transitions show the associated hysteretic effects (from Ref. 31).

Figures:

Fig. 1

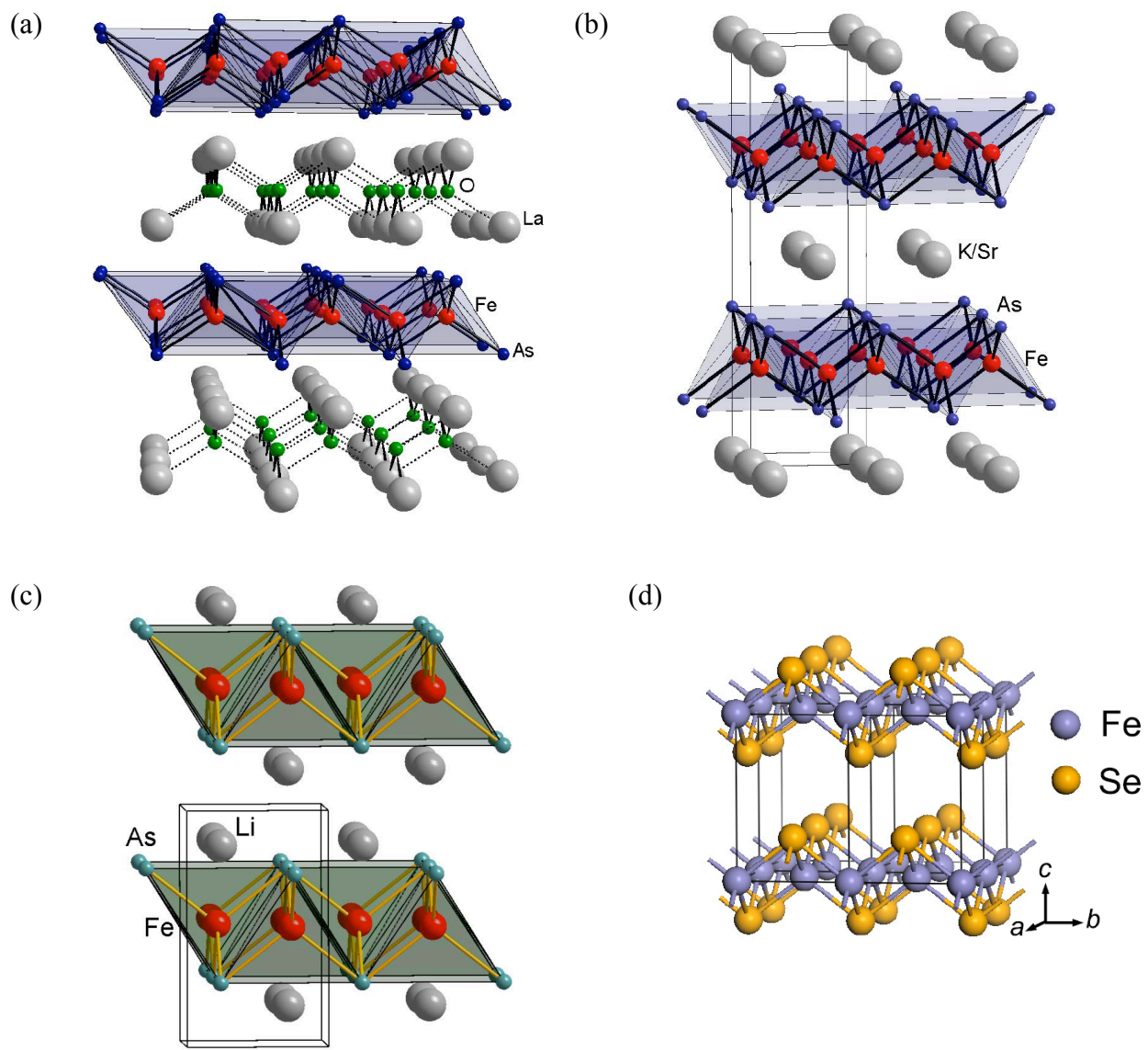

Fig. 2

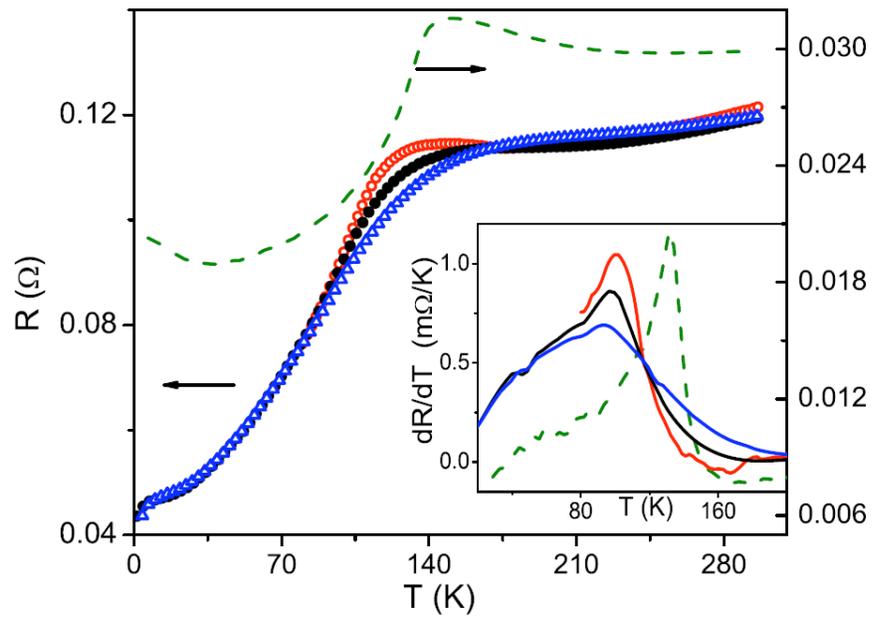

Fig. 3

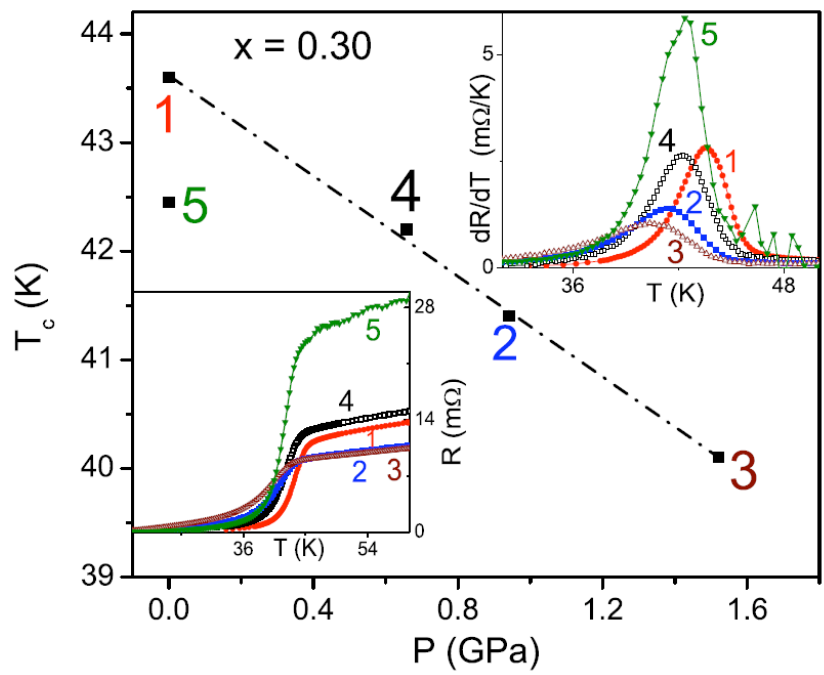

Fig. 4

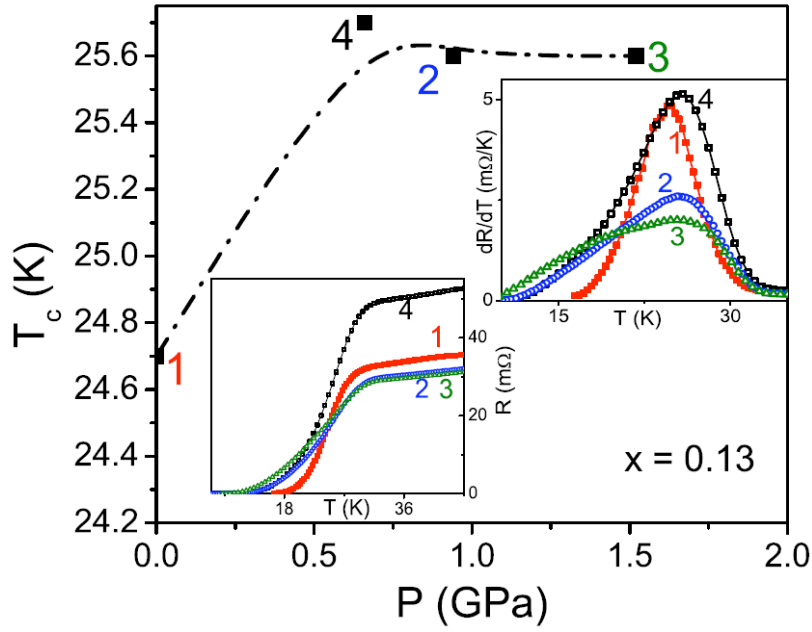

Fig. 5

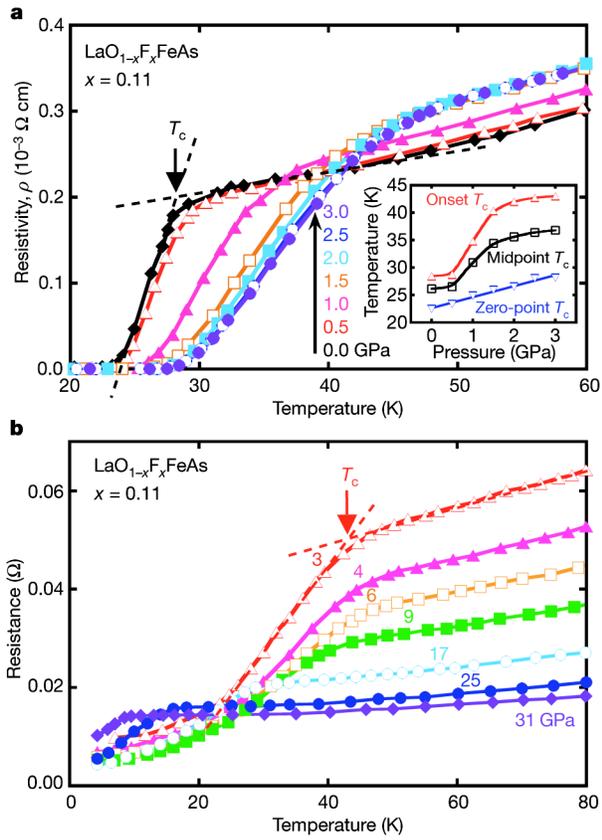

Fig. 6

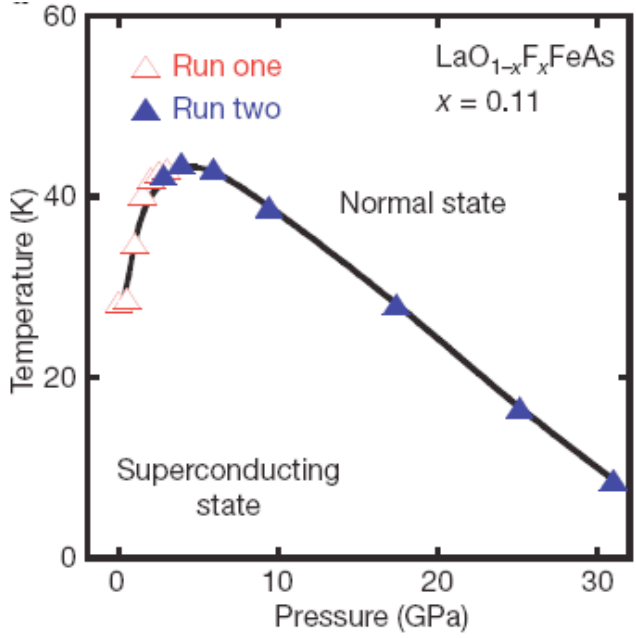

Fig. 7

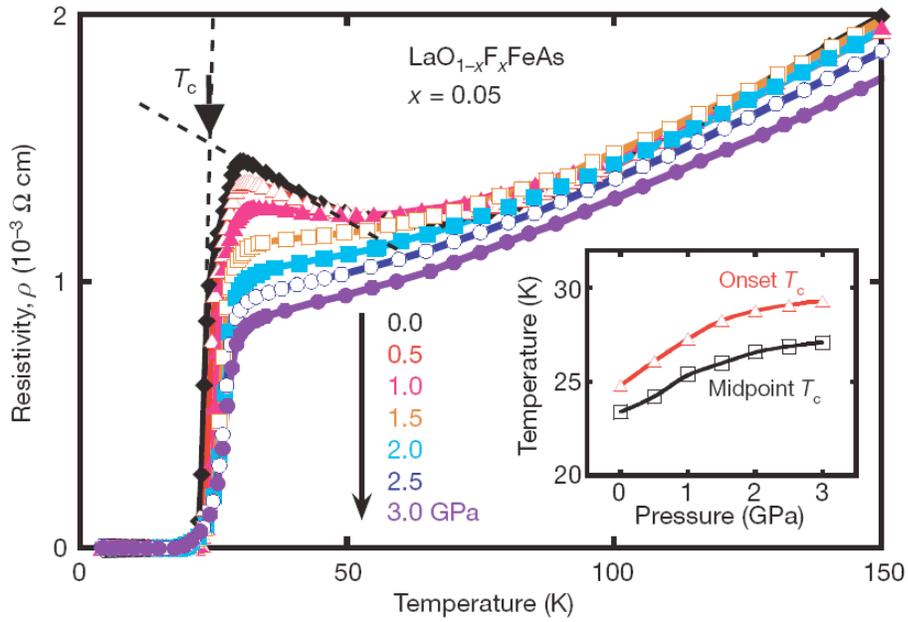

Fig. 8

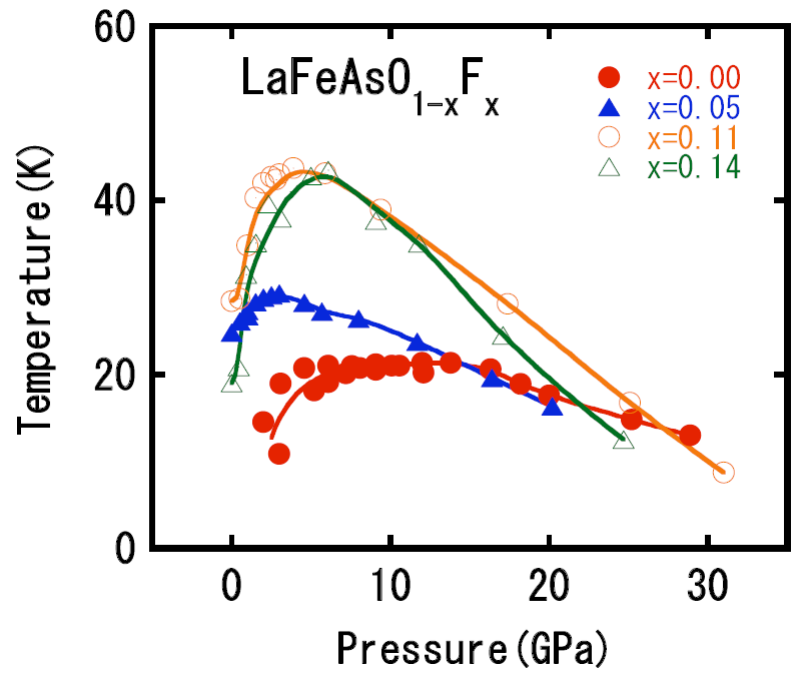

Fig. 9

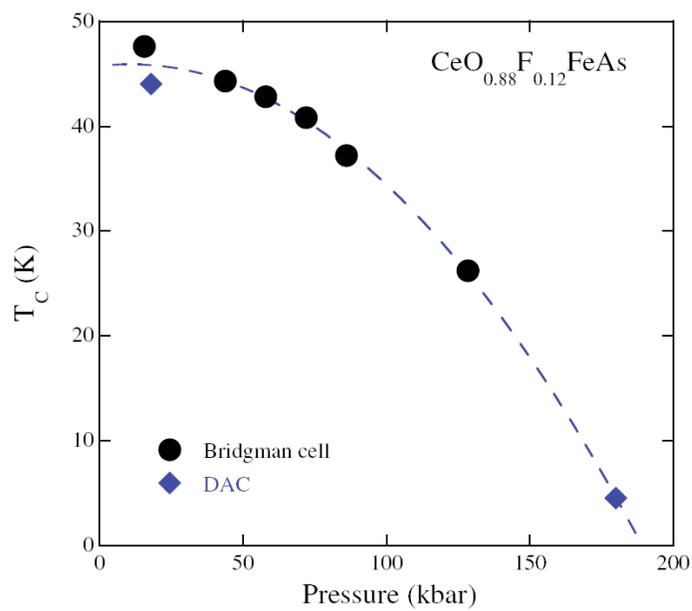

Fig. 10

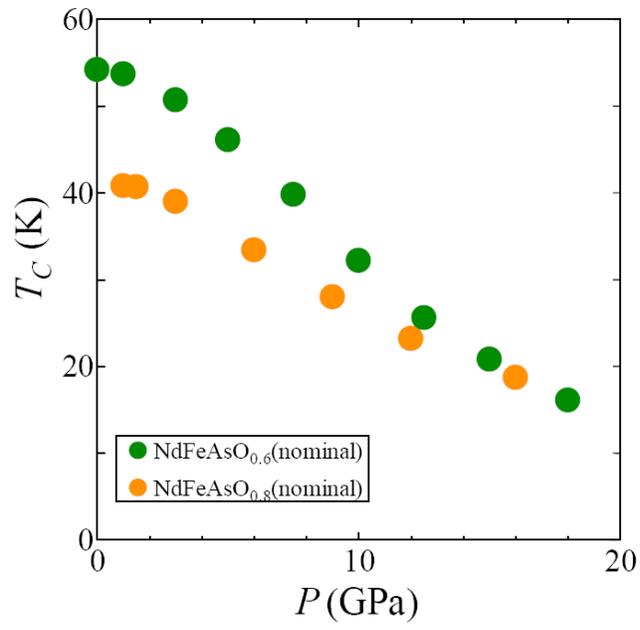

Fig. 11

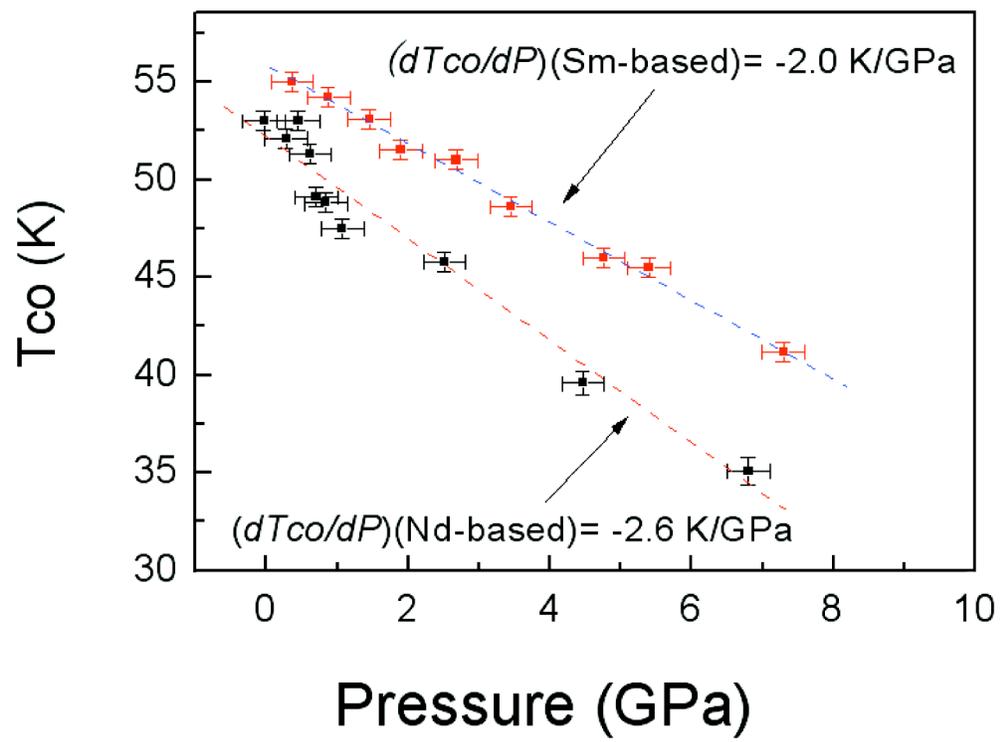

Fig. 12

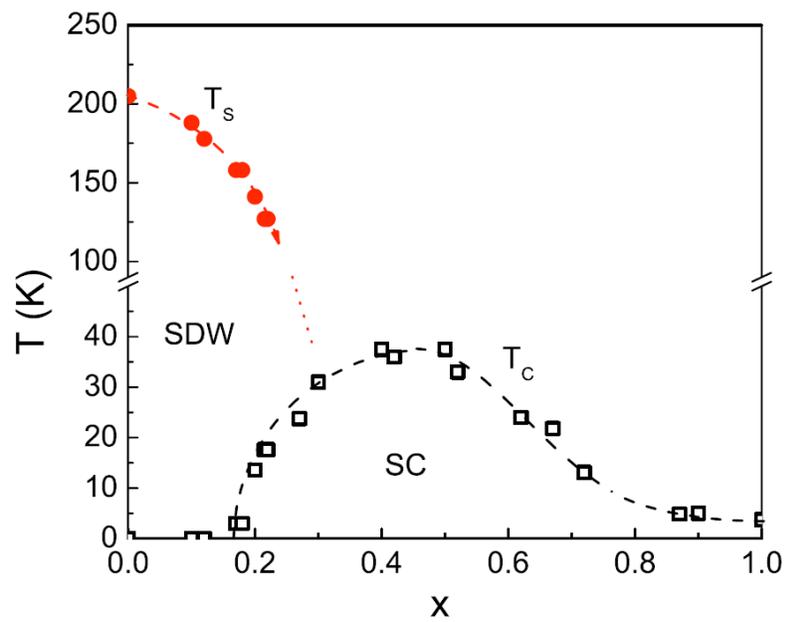

Fig. 13

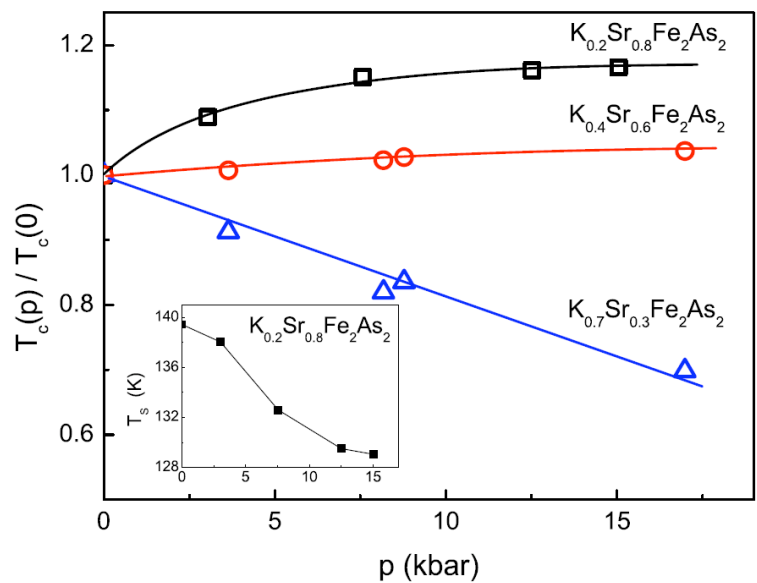

Fig. 14

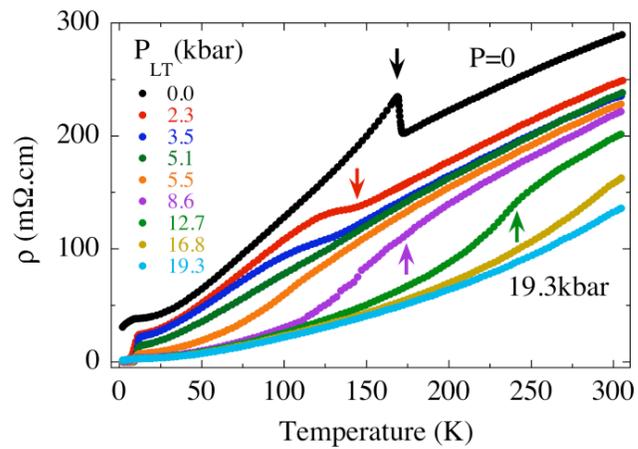

Fig. 15

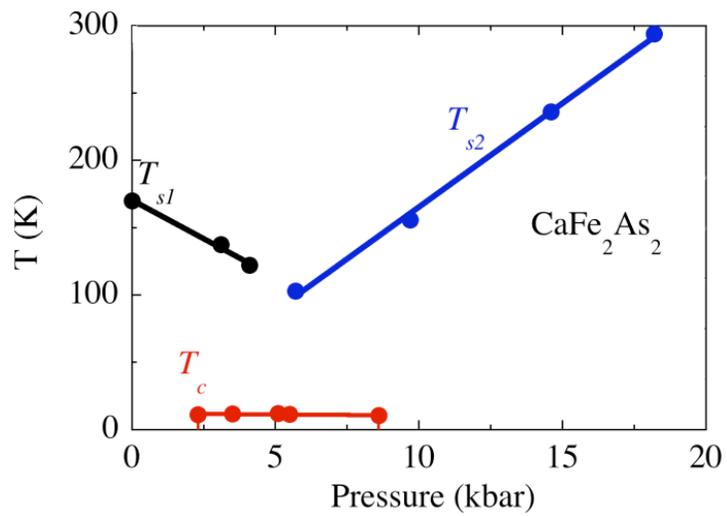

Fig. 16

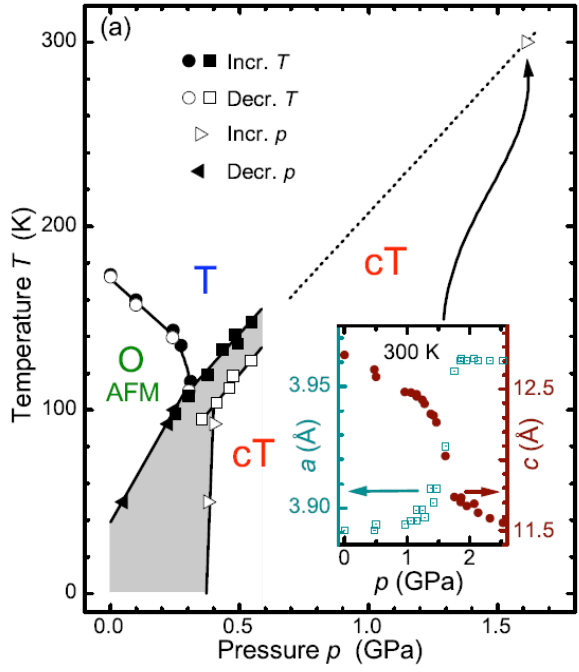

Fig. 17

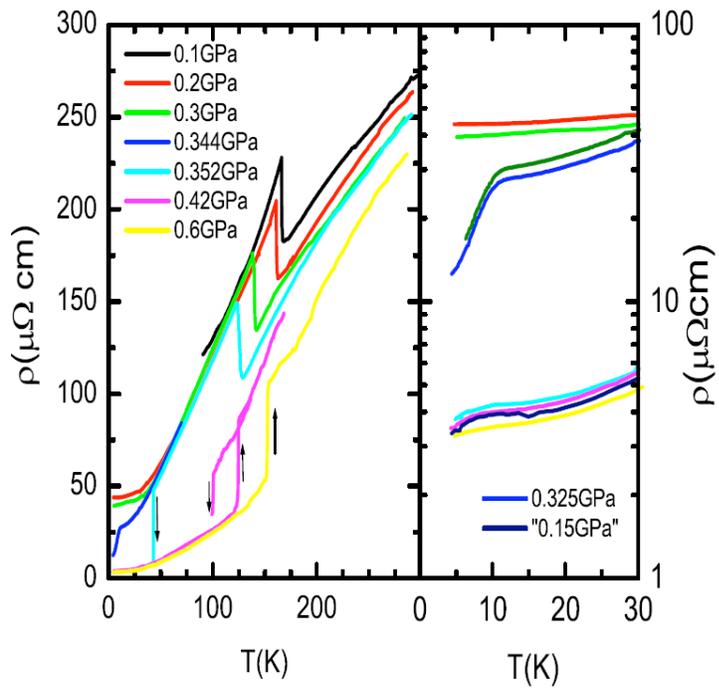